\documentclass[twocolumn,aps,showpacs,prl]{revtex4}
\usepackage{graphicx}
\usepackage{verbatim}
\begin{document}

\title{Fractality of Hofstadter Butterfly in Specific Heat Oscillation}

\author{L. P. Yang$^{1}$}\thanks{Present address: Institute for Theoretical Solid State Physics, IFW Dresden, 01069 Dresden, Germany}
\author{W. H. Xu$^{2}$}\email{wenhu@physics.rutgers.edu}
\author{M. P. Qin$^{3}$}
\author{T. Xiang$^{3,1}$}%\email{txiang@aphy.iphy.ac.cn}

\affiliation{$^1$Institute of Theoretical Physics, Chinese Academy
of Sciences, P.O. Box 2735, Beijing, 100190, China}

\affiliation{$^2$Department of Physics and Astronomy, Rutgers
University, 08854, USA}

\affiliation{$^3$Institute of Physics, Chinese Academy of Sciences,
P.O. Box 603, Beijing, 100190, China}

\begin{abstract}
We calculate thermodynamical properties of the Hofstadter model
using a recently developed quantum transfer matrix method. We find
intrinsic oscillation features in specific heat that manifest the
fractal structure of the Hofstadter butterfly. We also propose
experimental approaches which use specific heat as an access to
detect the Hofstadter butterfly.

\end{abstract}

\maketitle

The interplay between crystalline potential and magnetic field on a
two-dimensional electronic gas remains a nontrivial problem for
decades\cite{harper}. This issue provides a stage on which purely
mathematical concepts, i.e.,the irrationality of a real number,
interrupts our intuition of physical reality.
Hofstadter\cite{PhysRevB.14.2239} studied the energy spectrum of the
tight-binding limit of this problem, namely, the Hofstadter model.
He proposed a fractal topology for the spectrum(Hofstadter
butterfly), which reconciled the paradox raised by the
irrationality. The experimental verification of the Hofstadter
butterfly is challenging but some hints of the fractal structure
have been observed in microwave
measurements\cite{PhysRevLett.80.3232, PhysRevB.43.5192}, Hall
conductivity\cite{PhysRevLett.86.147} and magnetic transport
measurements\cite{PhysRevLett.62.1173} in analogous systems.

In this paper, we adopt a recently developed quantum transfer matrix
method\cite{Yang} to study thermodynamic properties of the
Hofstadter model. We focus on the behavior of internal energy and
specific heat as functions of magnetic field. As far as we know,
this is the first report that by theoretical method, the fractal
structure in the Hofstadter butterfly can be studied by computing
the specific heat in a magnetic field of a generic value. We also
briefly discuss the feasibility of experimental observations of
these features.

In a previous publication\cite{wenhu}, we had used the quantum
transfer matrix method to study the magnetic properties of
Hofstadter model. The advantage of this method lies in that it
directly computes the partition function of the model for arbitrary
$\phi$, where $\phi$ is the magnetic flux through a unit cell, then
the thermodynamic properties can be studied steadily. Conventional
theoretical methods, such as Bethe ansatz\cite{PhysRevLett.72.1890,
PhysRevLett.73.1134} and exact
diagonalization\cite{PhysRevLett.63.907, PhysRevLett.63.1657}, are
mostly applied to $\phi=p/q$ cases, where $p$ and $q$ are mutually
prime numbers, and $q$ is relatively small. Although detailed
information of energy spectrum and wavefunction can be obtained with
these methods, only limited cases of $\phi$ can be studied and most
discussion was focused on ground state properties. Besides, at
ground states, due to the fractality of the Hofstadter butterfly,
the smoothness of physical quantities as functions of magnetic
field, such as total energy, static magnetic susceptibility are
significantly diminished. However, within the quantum transfer
matrix formulation, the effect of finite temperature is embodied in
the partition function at the beginning, and the singularities due
to the fine fractality will be smeared out and the smoothness of
physical quantities can be recovered, which makes the comparison to
experimental results more straightforward.

Hofstadter model describes the dynamics of two-dimensional tight
binding electrons in a uniform magnetic
field\cite{PhysRevB.14.2239}. By applying Landau gauge, i.e.,
$A=H(0,x,0)$, the Hamiltonian is explicitly translationally
invariant along the $y$-direction. Fourier transformation along the
$y$-axis will then decouple the two-dimensional model $H$ into a
series summation of one-dimensional Hamiltonian $H_k$:
\begin{eqnarray} \label{eqn:ham}
H&=&\sum_{k} H_{k},\\
H_{k} & = & \sum_x \Big[ tc^{\dag}_{k,x+1}c_{k,x} +
tc^{\dag}_{k,x}c_{k,x+1}  \nonumber \\
&&  +2t\cos\left(2\pi x\phi-k\right) c^{\dag}_{k,x}c_{k,x} \Big] ,
\end{eqnarray}
where $k=2\pi n/N_y \, (n=0,1,...,N_y-1)$ are the quasimomenta and
$N_y$ is the lattice dimension along $y$ direction. $x$ is the
lattice coordinate of electrons along the $x$-axis. $\phi$ is the
magnetic flux through each plaquette, with magnetic flux quanta
$hc/e$ as unit.

$H_k$ does not generally have translational invariance along the
$x$-axis. But for rational $\phi=p/q$, periodicity can be recovered
by combining every $q$ cells to form a superlattice, and then the
problem can be solved by diagonalizing a $q\times q$ matrix for each
quasimomentum of the superlattice. Thus the full energy spectrum and
thermodynamic properties can be steadily obtained, yet apparently,
only up to relatively small $q$. To study cases with a generic
$\phi$, the quantum transfer matrix method starts from the partition
function $Z=\mathrm{Tr}[\exp(-\beta H)]$, which can be viewed as an
trace of the evolution operator along the imaginary time. Since the
trace naturally imposes a periodical boundary condition, a Fourier
transformation can be well defined along the imaginary time(the
inverse temperature), which is the key point leading us to the
transfer matrix representation and to significantly simplify the
calculation in Ref.[\onlinecite{wenhu, Yang}]. Given $k$, the
partition function of $H_k$ is defined by
\begin{equation}
Z_{k}=\mathrm{Tr} \exp(-\beta H_{k}),
\end{equation}
where $\beta = 1/ k_BT$. The partition function of the whole system
is simply a product of all $Z_k$s. By making use of the
translational invariance along the imaginary time, $Z_k$ can be
expressed as a product of $N_x$ $2\times2$ matrices. After
multiplying different $k$ components, we can obtain the partition
function of the system, from which one can calculate the free energy
by $F=-(1/\beta)\ln Z$, and other thermodynamic quantities such as
magnetic susceptibility and specific heat.

In Ref.[\onlinecite{wenhu}], the authors have discussed the effect
of lattice size on the numerical results. Accordingly, we choose
here $N_x=50000$ and $N_y=100$ to ensure the numerical accuracy as
well as computational efficiency for the temperature range in this
paper. For simplicity, we only consider the half-filling case, which
corresponds to a particle-hole symmetry and automatically sets
chemical potential $\mu$ to $0$.

First, we calculate the average internal energy as a function of
$\phi$ at $T=0.01$.
\begin{figure}[b]
\includegraphics[width=0.5\textwidth]{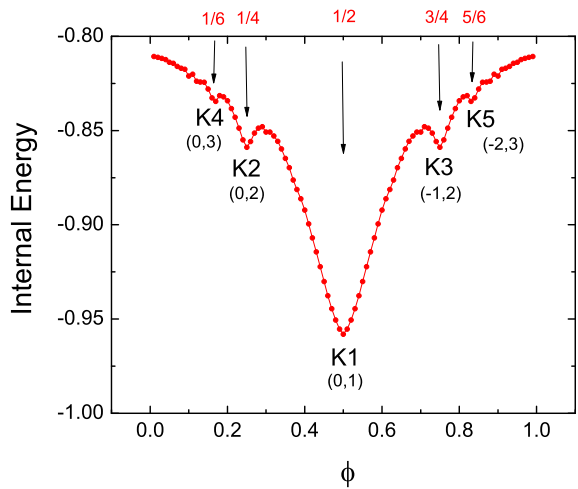}
\caption{(Color online)\label{fig:averE} The internal energy as a
function of $\phi$ for the Hofstadter model at half filling.
$T=0.01$. Some values of $\phi$(the red number above the top axis)
and the integers $(M, N)$ corresponding to local minima are marked.
}
\end{figure}
As shown in Fig.~\ref{fig:averE}, at the local minima of the
internal energy, the electron count $\nu$($=0.5$ for half-filling)
and $\phi$ satisfy the relation in (\ref{eqn:EvsPhi}), which was
given Ref.\cite{PhysRevLett.63.907}. These minima are {\it
cusp}-like.
\begin{equation}
\label{eqn:EvsPhi}
\nu=M + N\phi,~~~~~~M, N \in Z.
\end{equation}
The global minimum in Fig.~\ref{fig:averE} is consistent with the
conclusion that there is an global minimum of the average
energy\cite{PhysRevLett.63.1657, PhysRevB.41.9174} when
$\phi=\nu=1/2$, that is, each electron carries one flux quanta. We
have marked the values of $\phi$(the red number on the top axis)
located at distinguishable minima and the corresponding integers $M$
and $ N$ in Fig.~\ref{fig:averE}. At zero temperature, the average
energy will not be smooth almost everywhere because there are
infinite number of rational $\phi$s that satisfy (\ref{eqn:EvsPhi}).
But here the temperature will erase minor singularities and only
keep the significant ones.

\begin{figure}[b]
\includegraphics[width=0.5\textwidth]{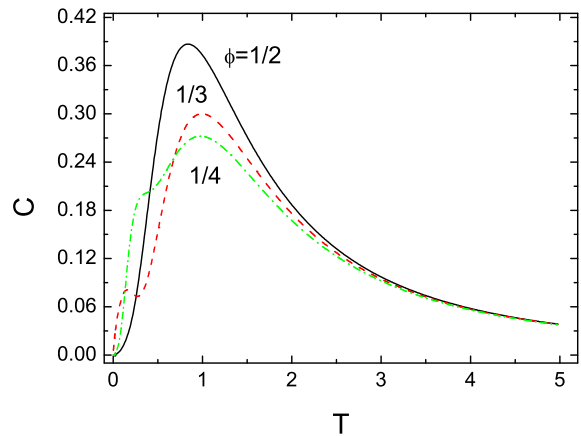}
\caption{(Color online)\label{fig:C-T} Temperature dependence of the
specific heat $C$ at half filling for $\phi=1/2,1/3,1/4$. }
\end{figure}

Then we compute the specific heat from the first order derivative of
the internal energy with respect to the temperature.
Fig.~\ref{fig:C-T} shows the specific heat $C$ as a function of
temperature $T$ for some special $\phi$s. The chosen three $\phi$s
belong to the pure cases in Hofstadter's
proposal\cite{PhysRevB.14.2239}, i.e.,$\phi=1/N$, or $1-1/N$ when
$N\geq2$. Under magnetic field of these values, the single Bloch
band in zero magnetic field is split into $N$ subbands. If $N$ is
odd, the central subband has a van-Hove singularity at the center
point of the energy spectrum($E=0$). If $N$ is even, the density of
states(DOS) goes to zero at $E=0$\cite{PhysRevLett.63.907}. When the
temperature is so high that the thermal fluctuations are comparable
to the energy difference between the lowest and the highest subband,
the subbands will not be able to manifest their internal fine
structures from specific heat. This can be observed from the high
temperature tail in Fig.~\ref{fig:C-T}.

The difference in the specific heat for various values of $\phi$
will emerge with the decreasing temperature. First, at low
temperature, the behavior of specific heat can tell the singularity
of DOS at the energy spectrum center point(the Fermi surface(FS) in
our half-filling case). In the regime near zero temperature, the
$\phi=1/2$ and $\phi=1/4$ curves are decreasing faster than that of
$\phi=1/3$. A closer observation indicates that $\phi=1/2$ and
$\phi=1/4$ decrease exponentially-like, while $\phi=1/3$ is
linear-like. This is because of the different behavior of DOS at the
spectrum center point\cite{PhysRevLett.63.907}. For $\phi=1/2$ and
$\phi=1/4$, the original single band in zero field splits up into
$2$ and $4$ bands. But the centermost two bands are not completely
separated by a gap, rather they ``kiss'' at the center point, where
DOS of both bands goes to zero. Therefore, we will expect a gap-like
behavior at low temperature, which shows up as an exponential-like
decrease in specific heat. For $\phi=1/3$, the Bloch band splits
into $3$ bands, and DOS of the center band is singular at the center
point. Thus we expect a much slower decrease near zero temperature,
which is also observed in Fig.~\ref{fig:C-T}.

In the intermediate temperature regime, the specific heat tells the
information about gaps and redistribution of DOS along the energy
spectrum. In Fig.~\ref{fig:C-T}, the curves of $\phi=1/3,1/4$ show
some similar minor hump structures, which is different from the case
of $\phi=1/2$. For $\phi=1/2$, two subbands touch at FS, where DOS
is zero, and there is no finite gap in the energy spectrum. Thus
there is only one major hump in specific heat curve. For both
$\phi=1/3$ and $\phi=1/4$, there is a finite gap lying above
FS\cite{PhysRevLett.63.907}, which separates the subband
on($\phi=1/3$) or near($\phi=1/4$) FS from the higher band, and
gives the extra minor hump in the specific heat curve.

\begin{figure}
\includegraphics[width=0.5\textwidth]{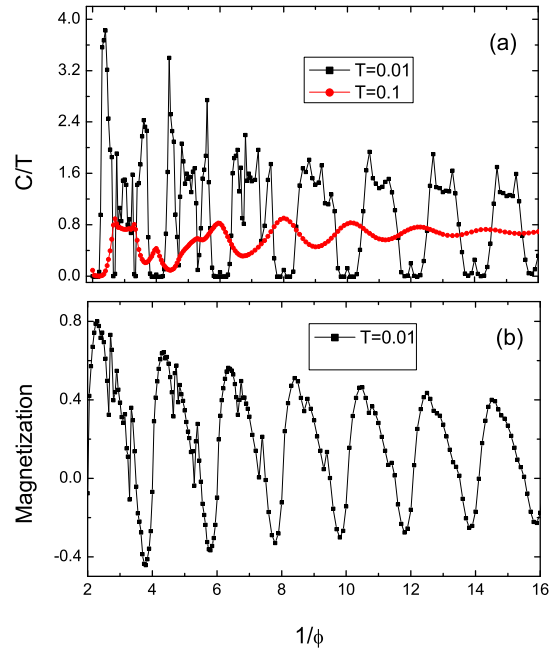}
\caption{(Color online)(a) The specific heat coefficient $C/T$ as a
function of $1/\phi$. Two different temperatures $T=0.01,0.1$ are
compared; (b) Magnetizationat at $T=0.01$ as a function of
$1/\phi$.} \label{fig:C-phi}
\end{figure}

Fig.~\ref{fig:C-phi}-(a) shows the specific heat coefficient($C/T$)
as a function of magnetic field at different temperatures, $T=0.1$
and $T=0.01$. The horizontal axis is chosen as $1/\phi$, so that the
conventional de Haas-van Alphen(dHvA) -like oscillations are shown
distinctly in the figure. The period $\triangle ( 1 / \phi )$ of
this oscillation is about $2$ in both cases, which is consistent
with that obtained from textbook formula\cite{solidstatephysics},
$\triangle ( 1 / \phi )= 4\pi^{2}/S_{F}$, where $S_F$ is the Fermi
volume. At half filling, $S_F = 2\pi^2$, thus $\triangle ( 1 / \phi
)=2$ in Fig.~\ref{fig:C-phi}-(a).

However, a more important observation is that subtle oscillations
emerge within the dHvA-type period with decreasing temperature and
strong field. The specific heat at $T=0.1$ displays a clean periodic
oscillation on the weaker field side(large $1/\phi$), while this
periodicity is disturbed in the stronger field regime(small
$1/\phi$). This becomes more explicit with lower temperature
$T=0.01$. In the first three periods, very sharp peaks and dips show
up, and they make a peculiar type of oscillations within the period.
Even for weaker field regime, some sharp structures are still
observable.

For the purpose of comparison, Fig.~\ref{fig:C-phi}-(b) shows the
magnetization oscillation with respect to the magnetic field.
Similar to the specific heat, the main envelope of oscillation is
the conventional dHvA oscillation. Besides, subtle structures emerge
within the dHvA period\cite{wenhu}. By comparing the results for
magnetization and specific heat, we find that the specific heat
oscillations are more distinct and drastic.

\begin{figure}
\includegraphics[width=0.5\textwidth]{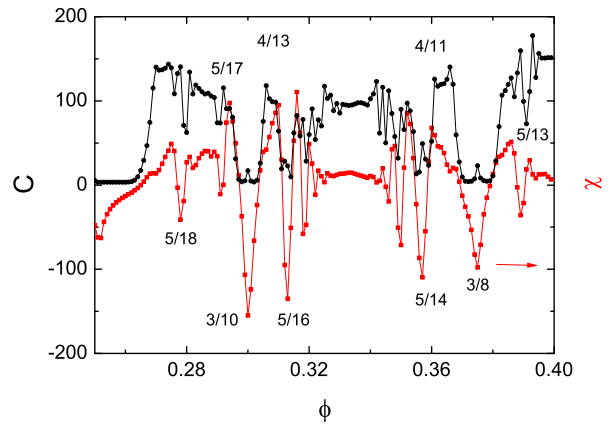}
\caption{(Color online) Specific heat coefficient(black line) and
magnetic susceptibility(red line) for the Hofstadter model at
half-filling. The numerical values of specific heat are 6000 times
larger than the original ones for comparison. The values of $\phi$s
marked corresponds to some local maxima and minima in both specific
heat and magnetic susceptibility. } \label{fig:C-phi_zoomin}
\end{figure}

To explore the information about the fractal structure of Hofstadter
butterfly from specific heat, we zoom in the first period in
Fig.\ref{fig:C-phi} and then have Fig.\ref{fig:C-phi_zoomin} for low
temperature specific heat $C$ and magnetic susceptibility $\chi$.
The numerical values of the specific heat are enlarged to 6000 times
of the original values for a clear comparison. Here $\phi$ is chosen
to be the horizontal axis. The consistency between $C$ and $\chi$ is
obvious if we compare the positions of local maxima and minima of
both quantities. In Fig.~\ref{fig:C-phi} some fractional values of
$\phi$s are marked where they are close to the local maxima and
minima. Applying Hofstadter's proposal of constructing the
butterfly\cite{PhysRevB.14.2239}, we can extract the structure of
energy spectrum at thess $\phi$s and understand why the extrema of
$C$ and $\chi$ are close to them. With Hofstadter's proposal, each
fractional $\phi$ can be decomposed to a set of more ``fundamental
'' fractions, or, ``local variable'' as in
Ref.\cite{PhysRevB.14.2239}, which then directly displays the
splitting of subbands in the energy spectrum. For example, for
$\phi=4/13$, the center local variable is $4/5$, which means there
is a cluster of $5$ subbands centered at FS. Consequently, a
van-Hove singularity of DOS shows up at FS and causes the local
maximum in $C$ and the strong paramagnetism(local maximum in
$\chi$)\cite{wenhu}, while for $\phi=3/8$, the center local variable
is $1/2$, thus there are two subbands lying above and below FS with
a zero DOS at FS and consequently in Fig.\ref{fig:C-phi_zoomin}, $C$
shows a small value(close to 0) around $\phi=3/8$ and $\chi$ is
strongly diamagnetic around $\phi=3/8$.

Therefore, by decreasing the temperature, fractal structures of the
Hofstadter butterfly manifest themselves by producing peculiar
oscillatory features within conventional dHvA period. This emergence
along with decreasing temperature is due to the fact that
temperature provides the only energy scale that sets up the
resolution of the spectrum. Temperature erases minor bands and gaps
that are smaller than the scale of temperature and restores the
smoothness of physical quantities. But fractal structures with an
energy scale larger than the temperature survive, and are able to
manifest themselves by displaying smoothened singularities in
thermodynamic quantities. Thus the subtler fractal structures of
Hofstadter butterfly can be probed by the measurement of the
specific heat at lower temperatures.

We propose to adopt the superconducting thin films(for example the
element Nb) with periodic arrays of pinning sites\cite{Harada} to
realize this temperature-dependent emergence of fractal structures
in specific heat. The artificial pinning centers hold great
potential. Just below the onset temperature of superconducting
transition, the electrons possess long mean free path. When the
interval between adjacent sites comes to the order of 100nm, the
experimentally accessible steady fields can enter the interesting
regime of $\phi$. The resulting effective lattice subjected to
perpendicular magnetic field is probably able to show the fractal
properties of the Hofstadter model. In addition, the purity
requirement of the sample is relaxed when considering the specific
heat measurement.

In summary, adopting the quantum transfer matrix method, we compute
the internal energy and specific heat of the Hofstadter model, and
for the first time we study the oscillation of the specific heat
with varying magnetic field as a signature of fractal structure of
the Hofstadter butterfly. In low field regime, the oscillation
period of specific heat is consistent with the conventional dHvA
oscillation. When the temperature is decreased, sharp peaks and dips
emerge in addition to the dHvA-type background. These peculiar
oscillatory behaviors are direct indications of DOS in the fractal
energy spectrum. We also suggest the possibility of making use of
superconducting films to detect this fractal structure by measuring
the specific heat.

Acknowledgement: We thank Shao-Jing Qin for helpful
discussions. One of us(Li-Ping Yang) is pleased to acknowledge
generous hospitality of Xiao Hu during her visit in Japan. This work
was supported by the NSF-China and the National Program for Basic
Research of MOST, China.

\bibliography{references}% Produces the bibliography via BibTeX.

\end{document}